\documentclass[12pt]{article}

\newcommand{\ie}   {{\em i.e.}}
\newcommand{\half}  {\frac{1}{2}}
\renewcommand{\bar}{\overline}

\newcommand{\VEV}[1]{\left\langle{#1}\right\rangle}
\newcommand{\ket}[1]{\vert\,{#1}\rangle}

\begin{document}

\begin{flushright}
{\small
SLAC--PUB--9689\\
JLAB-THY-03-31 \\
March 2003\\}
\end{flushright}

\vfill

\begin{center}
{{\bf\LARGE Light-Front Quantization of Gauge
Theories\footnote{\uppercase{T}his work is supported by the
Department of Energy under contract number DE--AC03--76SF00515.}}}

\bigskip
Stanley J. Brodsky\\
{\sl Stanford Linear Accelerator Center \\
Stanford University, Stanford, California 94309 \\
and \\
Thomas Jefferson National Accelerator Laboratory, Newport News, Virginia 23606\\
sjbth@slac.stanford.edu} \\
\medskip
\end{center}

\vfill

\begin{center}
Abstract \end{center}

Light-front wavefunctions provide a frame-independent
representation of hadrons in terms of their physical quark and
gluon degrees of freedom.  The light-front Hamiltonian formalism
provides new nonperturbative methods for obtaining the QCD
spectrum and eigensolutions, including resolvant methods,
variational techniques, and discretized light-front quantization.
A new method for quantizing gauge theories in light-cone gauge
using Dirac brackets to implement constraints is presented.  In
the case of the electroweak theory, this method of light-front
quantization leads to a unitary and renormalizable theory of
massive gauge particles, automatically incorporating the Lorentz
and 't Hooft conditions as well as the Goldstone boson equivalence
theorem. Spontaneous symmetry breaking is represented by the
appearance of zero modes of the Higgs field leaving the
light-front vacuum equal to the perturbative vacuum. I also
discuss an ``event amplitude generator" for automatically
computing renormalized amplitudes in perturbation theory. The
importance of final-state interactions for the interpretation of
diffraction, shadowing, and single-spin asymmetries in inclusive
reactions such as deep inelastic lepton-hadron scattering is
emphasized.

\vfill

\begin{center}
{\it Invited talk, presented at the\\
2002 International Workshop On Strong Coupling Gauge Theories
and\\
Effective Field Theories (SCGT 02)\\
Nagoya, Japan\\
 10--13 December 2002
 }\\
\end{center}

\vfill
\newpage

\section{Introduction}

Light-front wavefunctions are the amplitudes which interpolate
between hadrons and their quark and gluon degrees of freedom in
QCD.\cite{bro} For example, the eigensolution of a meson,
projected on the eigenstates $\{\ket{n} \}$ of the free
Hamiltonian $ H^{QCD}_{LC}(g = 0)$ at fixed light-front time $\tau
= t+z/c$ with the same global quantum numbers, has the expansion:
\begin{eqnarray}
\left\vert \Psi_M; P^+, {\vec P_\perp}, \lambda \right> &=&
\sum_{n \ge 2,\lambda_i} \int \Pi^{n}_{i=1} \frac{d^2k_{\perp i}
dx_i}{\sqrt{x_i} 16 \pi^3}\nonumber\\ &&
 \times 16 \pi^3 \delta\left(1- \sum^n_j x_j\right) \delta^{(2)}
\left(\sum^n_\ell \vec k_{\perp \ell}\right) \\[1ex]
&&\times \left\vert n; x_i P^+, x_i {\vec P_\perp} + {\vec
k_{\perp i}}, \lambda_i\right
> \psi_{n/M}(x_i,{\vec k_{\perp i}},\lambda_i)  .\nonumber
\end{eqnarray}
The set of light-front Fock state wavefunctions $\{\psi_{n/M}\}$
represents the ensemble of quark and gluon states possible when
the meson is intercepted at the light-front.  The light-front
momentum fractions $x_i = k^+_i/P^+_\pi = (k^0 + k^z_i)/(P^0+P^z)$
with $\sum^n_{i=1} x_i = 1$ and ${\vec k_{\perp i}}$ with
$\sum^n_{i=1} {\vec k_{\perp i}} = {\vec 0_\perp}$ represent the
relative momentum coordinates of the QCD constituents; the scalar
light-front wavefunctions $\psi_{n/p}(x_i,{\vec k_{\perp
i}},\lambda_i)$ are independent of the proton's momentum $P^+ =
P^0 + P^z$, and $P_\perp$. The physical transverse momenta are
${\vec p_{\perp i}} = x_i {\vec P_\perp} + {\vec k_{\perp i}}.$
The $\lambda_i$ label the light-front spin $S^z$ projections of
the quarks and gluons along the quantization $z$ direction.  The
spinors of the light-front formalism automatically incorporate the
Melosh-Wigner rotation.  Light-cone gauge $A^+=0$ is used to
eliminate unphysical gauge degrees of freedom.  The gluon
polarization vectors $\epsilon^\mu(k,\ \lambda = \pm 1)$ are
specified in light-cone gauge by the conditions $k \cdot \epsilon
= 0,\ \eta \cdot \epsilon = \epsilon^+ = 0.$ The quark and gluon
degrees of freedom are all physical; there are effectively no
ghost or negative metric states.

An important feature of the light-front formalism is that the
projection $J_z$ of the total angular momentum is kinematical and
conserved.  Each light-front Fock state component satisfies the
angular momentum sum rule: $ J^z = \sum^n_{i=1} S^z_i +
\sum^{n-1}_{j=1} l^z_j \ . $ The summation over $S^z_i$ represents
the contribution of the intrinsic spins of the $n$ Fock state
constituents.  The summation over orbital angular momenta
\begin{equation}
l^z_j = -{\mathrm i} \left(k^x_j\frac{\partial}{\partial k^y_j}
-k^y_j\frac{\partial}{\partial k^x_j}\right)
\end{equation}
derives from the $n-1$ relative momenta.  This excludes the
contribution to the orbital angular momentum due to the motion of
the center of mass, which is not an intrinsic property of the
hadron.  The light-front eigensolution corresponds to a spin $J$
particle in the hadron rest frame $P^+=P^-=M,\vec P_\perp=\vec 0$,
not the constituent rest frame $\sum_i \vec k_i = \vec 0$ since
$\sum_i k^z_i \ne P^z.$  The numerator structure of the
light-front wavefunctions is in large part determined by the
angular momentum constraints.  Thus wavefunctions generated by
perturbation theory\cite{Brodsky:2001ii,Ji:2003fw} can provide a
template for the numerator structure of nonperturbative
light-front wavefunctions.

Hadronic amplitudes can be computed by inserting a sum over a
complete sets of free Fock states for each external hadron, thus
representing the dynamics of each hadron as a convolution of its
light-front wavefunctions with the corresponding $n$-particle
irreducible quark-gluon matrix elements, summed over $n.$ For
example, in the case of spacelike form factors, the matrix
elements of local currents are given by a simple overlap of
light-front wavefunctions.  If one chooses  the frame with
$q^+=0$, then matrix elements of currents such as $j^+$ in
electroweak theory and have only diagonal matrix elements
$n^\prime =n.$ Thus once one has solved for the light-front
wavefunctions, one can compute hadron matrix elements of currents
between hadronic states of arbitrary momentum. Remarkably, quantum
fluctuations of the vacuum are absent if one uses light-front time
to quantize the system, so that matrix elements such as the
electromagnetic form factors only depend on the currents of the
constituents described by the light-cone wavefunctions.  As I
discuss below, the degrees of freedom associated with vacuum
phenomena such as spontaneous symmetry breaking in the Higgs model
have their counterpart in light-front $k^+ =0$ zero modes of the
fields.

Matrix elements of spacelike currents such as spacelike
electromagnetic form factors thus have an exact representation in
terms of simple overlaps of the light-front wavefunctions in
momentum space with the same $x_i$ and unchanged parton number
$n$.\cite{Drell:1970km,West:1970av,Brodsky:1980zm}  The Pauli form
factor and anomalous moment are spin-flip matrix elements of $j^+$
and thus connect states with $\Delta L_z =1.$\cite{Brodsky:1980zm}
Thus, these quantities are nonzero only if there is nonzero
orbital angular momentum of the quarks in the proton.   The Dirac
form factor is diagonal in $L_z$ and is typically dominated at
high $Q^2$ by highest states with the highest orbital angular
momentum. In the case of nuclear form factors, Fock states with
``hidden color" play an important role, particularly at large
momentum transfer.\cite{Brodsky:1983vf} The formulae for
electroweak current matrix elements of $j^+$ can be easily
extended to the $T^{++}$ coupling of gravitons.  In, fact, one can
show that the anomalous gravito-magnetic moment $B(0)$, analogous
to $F_2(0)$ in electromagnetic current interactions, vanishes
identically for any system, composite or
elementary.\cite{Brodsky:2001ii}  This important feature, which
follows in general from the equivalence
principle~\cite{Okun,Ji:1997nm,Teryaev:1999su}, is obeyed
explicitly in the light-front formalism.\cite{Brodsky:2001ii}

The light-front Fock representation is especially advantageous in
the study of exclusive $B$ decays.  For example, we can write down
an exact frame-independent representation of decay matrix elements
such as $B \to D \ell \bar \nu$ from the overlap of $n' = n$
parton conserving wavefunctions and the overlap of $n' = n-2$ from
the annihilation of a quark-antiquark pair in the initial
wavefunction.\cite{Brodsky:1999hn}  The off-diagonal $n+1
\rightarrow n-1$ contributions give a new perspective for the
physics of $B$-decays.  A semileptonic decay involves not only
matrix elements where a quark changes flavor, but also a
contribution where the leptonic pair is created from the
annihilation of a $q {\bar{q'}}$ pair within the Fock states of
the initial $B$ wavefunction.  The semileptonic decay thus can
occur from the annihilation of a nonvalence quark-antiquark pair
in the initial hadron.  Intrinsic charm $\ket{ b \bar u c \bar c}$
states of the $B$ meson, although small in probability, can play
an important role in its weak decays because they facilitate
CKM-favored weak decays.\cite{Brodsky:2001yt} The ``handbag"
contribution to the leading-twist off-forward parton distributions
measured in deeply virtual Compton scattering has a similar
light-front wavefunction representation as overlap integrals of
light-front wavefunctions.\cite{Brodsky:2000xy,Diehl:2000xz}

In the case of hadronic amplitudes involving a hard momentum
transfer $Q$, it is often possible to expand the quark-gluon
scattering amplitude as a function of ${k^2_\perp}/{Q^2}$. The
leading-twist contribution then can be computed from a
hard-scattering amplitude $T_H$ where the external quarks and
gluons emanating from each hadron can be taken as collinear.  The
convolution with the light-front wavefunction and integration
$\Pi_i d^2 k_{\perp i}$  over the relative transverse momentum
projects out only the $L_z=0$ component of the light-front
wavefunctions.  This leads to hadron spin selection rules such as
hadron helicity conservation.\cite{Brodsky:1981kj}  Furthermore,
only the minimum number of quark and gluon quanta contribute at
leading order in $1/Q^2.$ The nominal scaling of  hard hadron
scattering amplitudes at leading twist then obeys dimensional
counting rules.\cite{Brodsky:1973kr,Matveev:ra,Brodsky:1974vy}
Recently these rules have been derived to all orders in the gauge
coupling in conformal QCD and large $N_C$ using gauge/string
duality.\cite{Polchinski:2001tt} There is also evidence from
hadronic $\tau$ decays that the QCD coupling approaches an
infrared fixed-point at low scales.\cite{Brodsky:2002nb} This may
explain the empirical success of conformal approximations to QCD.
The distribution amplitudes $\phi(x_i,Q)$ which appear in
factorization formulae for hard exclusive processes are the
valence LF Fock wavefunctions integrated over the relative
transverse momenta up to the resolution scale
$Q$.\cite{Lepage:1980fj}  These quantities specify how a hadron
shares its longitudinal momentum among its valence quarks; they
control virtually all exclusive processes involving a hard scale
$Q$, including form factors, Compton scattering, semi-exclusive
processes,\cite{Brodsky:1998sr} and photoproduction at large
momentum transfer, as well as the decay of a heavy hadron into
specific final states.\cite{Beneke:1999br,Keum:2000ph}

The quark and gluon probability distributions $q_i(x,Q)$ and
$g(x,Q)$ of a hadron can be computed from the absolute squares of
the light-front wavefunctions, integrated over the transverse
momentum.  All helicity distributions are thus encoded in terms of
the light-front wavefunctions.  The DGLAP evolution of the
structure functions can be derived from the high $k_\perp$
properties of the light-front wavefunctions.  Thus given the
light-front wavefunctions, one can compute\cite{Lepage:1980fj} all
of the leading twist helicity and transversity distributions
measured in polarized deep inelastic lepton scattering.  Similarly,
the transversity distributions and off-diagonal helicity
convolutions are defined as a density matrix of the light-front
wavefunctions.

However, it is not true that the leading-twist structure functions
$F_i(x,Q^2)$  measured in deep inelastic lepton scattering are
identical to the quark and gluon distributions.  It is usually
assumed, following the parton model,  that the $F_2$ structure
function measured in neutral current deep inelastic lepton
scattering is at leading order in $1/Q^2$ simply $F_2(x,Q^2)
=\sum_q  e^2_q  x q(x,Q^2)$, where $x = x_{bj} = Q^2/2 p\cdot q$
and $q(x,Q)$ can be computed from the absolute square of the
proton's light-front wavefunction. Recent work by Hoyer, Marchal,
Peigne, Sannino, and myself shows that this standard
identification is wrong.\cite{Brodsky:2002ue} In fact, one cannot
neglect the Wilson line integral between currents in the current
correlator even in light-cone gauge.  In the case of light-cone
gauge, the Wilson line involves the transverse gluon field
$A_\perp$ not $A^+.$\cite{Belitsky:2002sm} Gluon exchange between
the fast, outgoing partons and the target spectators affects the
leading-twist structure functions in a profound way.  The
final-state interactions lead to the Bjorken-scaling diffractive
component $\gamma^* p \to p X$ of deep inelastic scattering.  The
diffractive scattering of the fast outgoing quarks on spectators
in the target in turn causes shadowing in the DIS cross section.
Thus the depletion of the nuclear structure functions is not
intrinsic to the wave function of the nucleus, but is a coherent
effect arising from the destructive interference of diffractive
channels induced by final-state interactions.    Similarly, the
effective Pomeron distribution of a hadron is not derived from its
light-front wavefunction and thus is not a universal property.
Many properties involving parton transverse momentum are also
affected by the Wilson line.\cite{Goeke:2003az}

Measurements from the HERMES and SMC collaborations show a
remarkably large single-spin asymmetry in semi-inclusive pion
leptoproduction $\gamma^*(q) p \to \pi X$ when the proton is
polarized normal to the photon-to-pion production plane.
Hwang, Schmidt, and I~\cite{Brodsky:2002cx} have shown that
final-state interactions from gluon exchange between the outgoing
quark and the target spectator system lead to single-spin
asymmetries in deep inelastic lepton-proton scattering at leading
twist in perturbative QCD; {\it i.e.}, the rescattering
corrections are not power-law suppressed at large photon
virtuality $Q^2$ at fixed $x_{bj}$.  The existence of such
single-spin asymmetries requires a phase difference between two
amplitudes coupling the proton target with $J^z_p = \pm
\frac{1}{2}$ to the same final-state, the same amplitudes which
are necessary to produce a nonzero proton anomalous magnetic
moment.  The single-spin asymmetry which arises from such
final-state interactions does not factorize into a product of
distribution function and fragmentation function, and it is not
related to the transversity distribution $\delta q(x,Q)$ which
correlates transversely polarized quarks with the spin of the
transversely polarized target nucleon.  In general all measures of
quark and gluon transverse momentum require consideration of
final-state interactions as incorporated in the Wilson line.

These effects
highlight the unexpected importance of final- and initial-state
interactions in QCD observables---they lead to leading-twist single-spin
asymmetries, diffraction, and nuclear shadowing, phenomena not included
in the light-front wavefunctions of the target.  Alternatively, as
discussed by Belitsky, Ji, and Yuan,\cite{Belitsky:2002sm} one can
augment the light-front wavefunctions by including the phases induced by
initial and final state interactions. Such wavefunctions correspond to
solving the light-front bound state equation in an external field.

\section{The Light-Front Quantization of QCD}

In Dirac's ``Front Form"\cite{Dirac:cp}, the generator of
light-front time translations is $P^- = i\frac{ \partial}{\partial
\tau}.$ Boundary conditions are set on the transverse plane
labelled by $x_\perp$ and $x^- = z-ct$.  Given the Lagrangian of a
quantum field theory, $P^-$ can be constructed as an operator on
the Fock basis, the eigenstates of the free theory.  Since each
particle in the Fock basis is on its mass shell, $k^- \equiv
k^0-k^3 = \frac{k^2_\perp + m^2}{k^+},$ and its energy $k^0 =\half
( k^+ + k^-) $ is positive, only particles with positive momenta
$k^+ \equiv k^0 + k^3 \ge 0$ can occur in the Fock basis.  Since
the total plus momentum $P^+ = \sum_n k^+_n$ is conserved, the
light-cone vacuum cannot have any particle content.

The Heisenberg equation on the light-front is
\begin{equation}
H_{LC} \ket{\Psi} = M^2 \ket{\Psi}\ .
\end{equation}
The operator $H_{LC} = P^+ P^- - P^2_\perp,$ the ``light-cone
Hamiltonian", is frame-independent.  This can in principle be
solved by diagonalizing the matrix $\VEV{n|H_{LC}|m}$ on the free
Fock basis:~\cite{Brodsky:1997de}
\begin{equation}
\sum_m \VEV{n|H_{LC}|m}\VEV{m|\psi} = M^2 \VEV{n|\Psi}\ .
\end{equation}
The eigenvalues $\{M^2\}$ of $H_{LC}=H^{0}_{LC} + V_{LC}$ give the
squared invariant masses of the bound and continuum spectrum of
the theory.  The light-front Fock space is the eigenstates of the
free light-front Hamiltonian; \ie, it is a Hilbert space of
non-interacting quarks and gluons, each of which satisfy $k^2 =
m^2$ and $k^- = \frac{m^2 + k^2_\perp}{k^+} \ge 0.$  The
projections $\{\VEV{n|\Psi}\}$ of the eigensolution on the
$n$-particle Fock states provide the light-front wavefunctions.
Thus solving a quantum field theory is equivalent to solving a
coupled many-body quantum mechanical problem:
\begin{equation}
\left[M^2 - \sum_{i=1}^n\frac{m^2 + k^2_\perp}{x_i}\right] \psi_n
= \sum_{n'}\int \VEV{n|V_{LC}|n'} \psi_{n'}\end{equation}
where the convolution and sum is understood over the Fock number,
transverse momenta, plus momenta, and helicity of the intermediate
states.  Light-front wavefunctions are also related to
momentum-space Bethe-Salpeter wavefunctions by integrating over
the relative momenta $k^- = k^0 - k^z$ since this projects out the
dynamics at $x^+ =0.$

A review of the development of light-front quantization of QCD and
other quantum field theories is given in the
references.\cite{Brodsky:1997de} The light-front quantization of
gauge theory can be most conveniently carried out in the
light-cone gauge $A^+ = A^0 + A^z = 0$.  In this gauge the $A^-$
field becomes a dependent degree of freedom, and it can be
eliminated from the Hamiltonian in favor of a set of specific
instantaneous light-front time interactions.  In fact in
$QCD(1+1)$ theory, this instantaneous interaction provides the
confining linear $x^-$ interaction between quarks.  In $3+1$
dimensions, the transverse field $A^\perp$ propagates massless
spin-one gluon quanta with polarization
vectors\cite{Lepage:1980fj} which satisfy both the gauge condition
$\epsilon^+_\lambda = 0$ and the Lorentz condition $k\cdot
\epsilon= 0$.

Prem Srivastava and I\cite{Srivastava:2000cf} have  presented a
new systematic study of light-front-quantized gauge theory in
light-cone gauge using a Dyson-Wick S-matrix expansion based on
light-front-time-ordered products.  The Dirac bracket method is
used to identify the independent field degrees of
freedom.\cite{Ditman}   In our analysis one imposes the light-cone
gauge condition as a linear constraint using a Lagrange
multiplier, rather than a quadratic form.  We then find that the
LF-quantized free gauge theory simultaneously satisfies the
covariant gauge condition $\partial\cdot A=0$ as an operator
condition as well as the LC gauge condition.  The gluon propagator
has the form
\begin{equation}
\VEV{0|\,T({A^{a}}_{\mu}(x){A^{b}}_{\nu}(0))\,|0}
=\frac{{i\delta^{ab}}}{{(2\pi)^{4}}} \int d^{4}k \;e^{-ik\cdot x}
\; \; \frac{D_{\mu\nu}(k)}{{k^{2}+i\epsilon}}
\end{equation}
where we have defined
\begin{equation}
D_{\mu\nu}(k)= D_{\nu\mu}(k)= -g_{\mu\nu} + \frac
{n_{\mu}k_{\nu}+n_{\nu}k_{\mu}}{(n\cdot k)} - \frac {k^{2}}
{(n\cdot k)^{2}} \, n_{\mu}n_{\nu}.
\end{equation}
Here $n_{\mu}$ is a null four-vector, gauge direction, whose
components are chosen to be $\, n_{\mu}={\delta_{\mu}}^{+}$, $\,
n^{\mu}={\delta^{\mu}}_{-}$.  Note also
\begin{eqnarray}
D_{\mu\lambda}(k) {D^{\lambda}}_{\nu}(k)=
D_{\mu\perp}(k) {D^{\perp}}_{\nu}(k)&=& - D_{\mu\nu}(k),  \\
 k^{\mu}D_{\mu\nu}(k)=0,  n^{\mu}D_{\mu\nu}(k)&\equiv&
D_{-\nu}(k)=0, \nonumber \\ D_{\lambda\mu}(q) \,D^{\mu\nu}(k)\,
D_{\nu\rho}(q') &=& -D_{\lambda\mu}(q)D^{\mu\rho}(q').\nonumber
\end{eqnarray}
The gauge field propagator $\,\,i\,D_{\mu\nu}(k)/
(k^{2}+i\epsilon)\,$ is transverse not only to the gauge direction
$n_{\mu}$ but also to $k_{\mu}$, {\em i.e.}, it is {\it
doubly-transverse}.  Thus $D$ represents the polarization sum over
physical propagating modes.  The last term proportional to $n_\mu
n_\nu$ in the gauge propagator does not  appear in the usual
formulations of light-cone gauge.  However, in tree graph
calculations it cancels against instantaneous gluon exchange
contributions.

The remarkable properties of (the projector) $D_{\nu\mu}$ greatly
simplifies the computations of loop amplitudes.  For example, the
coupling of gluons to propagators carrying high momenta is
automatic.  In the case of tree graphs, the term proportional to
$n_{\mu}n_{\nu}$ cancels against the instantaneous gluon exchange
term.  However, in the case of loop diagrams, the separation needs
to be maintained so that one can identify the correct
one-particle-irreducible contributions.   The absence of collinear
divergences in irreducible diagrams in the light-cone gauge
greatly simplifies the leading-twist factorization of soft and
hard gluonic corrections in high momentum transfer inclusive and
exclusive reactions\cite{Lepage:1980fj} since the numerators
associated with the gluon coupling only have transverse
components.

The interaction Hamiltonian of QCD in light-cone gauge can be
derived by systematically applying the Dirac bracket method to
identify the independent fields.\cite{Srivastava:2000cf}   It
contains the usual Dirac interactions between the quarks and
gluons, the three-point and four-point gluon non-Abelian
interactions plus instantaneous light-front-time gluon exchange
and quark exchange contributions
\begin{eqnarray}
{\mathcal H}_{int}&=&
  -g \,{{\bar\psi}}^{i}
\gamma^{\mu}{A_{\mu}}^{ij}{{\psi}}^{j}   \nonumber \\
&& +\frac{g}{2}\, f^{abc} \,(\partial_{\mu}{A^{a}}_{\nu}-
\partial_{\nu}{A^{a}}_{\mu}) A^{b\mu} A^{c\nu} \nonumber \\
&& +\frac {g^2}{4}\,
f^{abc}f^{ade} {A_{b\mu}} {A^{d\mu}} A_{c\nu} A^{e\nu} \nonumber \\
&& - \frac{g^{2}}{ 2}\,\, {{\bar\psi}}^{i} \gamma^{+}
\,(\gamma^{\perp'}{A_{\perp'}})^{ij}\,\frac{1}{i\partial_{-}} \,
(\gamma^{\perp} {A_{\perp}})^{jk}\,{\psi}^{k} \nonumber \\
&& -\frac{g^{2}}{ 2}\,{j^{+}}_{a}\, \frac
{1}{(\partial_{-})^{2}}\, {j^{+}}_{a}
\end{eqnarray}
where
\begin{equation}
{j^{+}}_{a}={{\bar\psi}}^{i} \gamma^{+} (
{t_{a}})^{ij}{{\psi}}^{j} + f_{abc} (\partial_{-} A_{b\mu})
A^{c\mu} \ .
\end{equation}

The renormalization constants in the non-Abelian theory have been
shown \cite{Srivastava:2000cf} to satisfy the identity $Z_1=Z_3$
at one-loop order, as expected in a theory with only physical
gauge degrees of freedom.  The renormalization factors in the
light-cone gauge are independent of the reference direction
$n^\mu$.   The QCD $\beta$ function computed in the noncovariant
LC gauge agrees with the conventional theory
result.\cite{gross,polit}  Dimensional regularization and the
Mandelstam-Leibbrandt
prescription\cite{Mandelstam:1982cb,Leibbrandt:1987qv,Bassetto:1984dq}
for LC gauge were used to define the Feynman loop
integrations.\cite{Bassetto:1996ph}  There are no Faddeev-Popov or
Gupta-Bleuler ghost terms.

The running coupling constant and the QCD $\beta$ function have
also been computed at one loop in the doubly-transverse light-cone
gauge.\cite{Srivastava:2000cf}   It is also possible to
effectively quantize QCD using light-front methods in covariant
Feynman gauge.\cite{Srivastava:2000gi}  It is well-known that the
light-cone gauge itself is not completely defined until one
specifies a prescription for the poles of the gauge propagator at
$n \cdot k= 0.$ The Mandelstam-Liebbrandt prescription has the
advantage of preserving causality and analyticity, as well as
leading to proofs of the renormalizability and unitarity of
Yang-Mills theories.\cite{Bassetto:1991ue}  The ghosts which
appear in association with the Mandelstam-Liebbrandt prescription
from the single poles have vanishing residue in absorptive parts,
and thus do not disturb the unitarity of the theory.

A remarkable advantage of light-front quantization is that the
vacuum state $\ket{0}$ of the full QCD Hamiltonian evidently
coincides with the free vacuum.  The light-front vacuum is
effectively trivial if the interaction Hamiltonian applied to the
perturbative vacuum is zero.  Note that all particles in the
Hilbert space have positive energy $k^0 = \frac{1}{2}(k^+ + k^-)$,
and thus positive light-front $k^\pm$.  Since the plus momenta
$\sum k^+_i$ is conserved by the interactions, the perturbative
vacuum can only couple to states with particles in which all
$k^+_i$ = 0; \ie, so called zero-mode states.   Bassetto and
collaborators\cite{Bassetto:1999tm} have shown that the
computation of the spectrum of $QCD(1+1)$ in equal time
quantization requires constructing the full spectrum of non
perturbative contributions (instantons).  In contrast, in the
light-front quantization of gauge theory, where the $k^+ = 0 $
singularity of the instantaneous interaction is defined by a
simple infrared regularization, one obtains the correct spectrum
of $QCD(1+1)$ without any need for vacuum-related contributions.
Zero modes of auxiliary fields are necessary to distinguish the
theta-vacua of massless QED(1+1)
\cite{Yamawaki:1998cy,McCartor:2000yy,Srivastava:1999et}, or to
represent a theory in the presence of static external boundary
conditions or other constraints.  Zero-modes provide the
light-front representation of spontaneous symmetry breaking in
scalar theories.\cite{Pinsky:1994yi}

\section{Light-Front Quantization of the Standard Model}

Prem Srivastava and I have also shown how light-front quantization
can be applied to the Glashow, Weinberg and Salam (GWS) model of
electroweak interactions based on the nonabelian gauge group
$SU(2)_{W}\times U(1)_{Y}$.\cite{gws}    This theory contains a
nonabelian Higgs sector which triggers spontaneous symmetry
breaking (SSB).  A convenient way of implementing SSB and the
(tree level) Higgs mechanism in the {\it front form} theory  was
developed earlier by Srivastava.\cite{pre4,pre5,pre6}  One
separates the quantum fluctuation fields from the corresponding
{\it dynamical bosonic condensate } (or
zero-longitudinal-momentum-mode) variables, {\it before} applying
the Dirac procedure in order to construct the Hamiltonian
formulation.  The canonical quantization of LC gauge GWS
electroweak theory in the {\it front form} can  be derived by
using the Dirac procedure to construct a self-consistent LF
Hamiltonian theory.  This leads to an attractive new formulation
of the Standard Model of the strong and electroweak interactions
which does not break the physical vacuum and has well-controlled
ultraviolet behavior.  The only ghosts which appear in the
formalism are the $n \cdot k = 0$ modes of the gauge field
associated with regulating the light-cone gauge prescription.  The
massive gauge field propagator has good asymptotic behavior in
accordance with a renormalizable theory, and the massive would-be
Goldstone fields can be taken as physical degrees of freedom.

For example, consider the Abelian Higgs model.  The interaction
Lagrangian is
\begin{equation} {\mathcal L}= -\frac{1}{4} F_{\mu
\nu}F^{\mu \nu}+ \vert D_\mu \phi\vert^2 -V(\phi^\dagger \phi)
\end{equation}
 where
\begin{equation}
D_\mu = \partial_\mu + i e A_\mu,\end{equation}  and
\begin{equation}V(\phi)= \mu^2 \phi^\dagger \phi + \lambda(\phi^\dagger
\phi)^2,\end{equation}
 with $\mu^2 < 0, \lambda >0.$ The complex
scalar field $\phi$ is decomposed as \begin{equation}\phi(x)=
\frac{1}{\sqrt 2} v + \varphi(x) = \frac{1}{\sqrt 2}[ v + h(x) + i
\eta(x)]\end{equation}
 where $v$ is the $k^+=0$ zero mode
determined by the minimum of the potential: $v^2 =
-\frac{\mu^2}{\lambda}$, $h(x)$ is the dynamical Higgs field, and
$\eta(x)$ is the Nambu-Goldstone field.  The quantization
procedure determines $\partial \cdot A = MR$, the 't Hooft
condition.  One can now eliminate the zero mode component of the
Higgs field $v$ which gives masses for the fundamental quantized
fields.   The $A_\perp$ field then has mass $M=e v$ and the Higgs
field acquires mass $m^2_h = 2 \lambda v^2 = - 2 \mu^2.$

A new aspect of LF quantization, is that the third polarization of
the quantized massive vector field $A^\mu$ with four momentum
$k^\mu$ has the form $E^{(3)}_\mu = {n_\mu M / n \cdot k}$.  Since
$n^2 = 0$, this non-transverse polarization vector has zero norm.
However, when one includes the constrained interactions of the
Goldstone particle, the effective longitudinal polarization vector
of a produced vector particle is $E^{(3)}_{\rm eff \, \mu}=
E^{(3)}_\mu - { k_\mu \,k \cdot E^{(3)} / k^2}$ which is identical
to the usual polarization vector of a massive vector with norm
$E^{(3)}_{\rm eff }\cdot E^{(3)}_{\rm eff }= -1$.  Thus, unlike
the conventional quantization of the Standard Model, the Goldstone
particle only provides part of the physical longitudinal mode of
the electroweak particles.

In the LC gauge LF framework, the free massive gauge fields in the
electroweak theory satisfy simultaneously the 't Hooft conditions
as an operator equation.  The sum over the three physical
polarizations is given by $K_{\mu\nu}$
\begin{eqnarray} K_{\mu\nu}(k)&=&\,\sum_{(\alpha)}
E^{(\alpha)}_{\mu}E^{(\alpha)}_{\nu} =\,D_{\mu\nu}(k)+
\frac{M^{2}}{(k^{+})^{2}}\, n_{\mu} n_{\nu}  \\
&=&-g_{\mu\nu} + \frac {n_{\mu}k_{\nu}+n_{\nu}k_{\mu}}{(n\cdot k)}
- \frac {(k^{2}-M^{2})} {(n\cdot k)^{2}} \,
n_{\mu}n_{\nu}\nonumber
\end{eqnarray} which satisfies:  $ k^{\mu}\,K_{\mu\nu}(k)=
(M^{2}/k^{+})\, n_{\nu}$ and $\,\, k^{\mu}\,
k^{\nu}\,K_{\mu\nu}(k)= M^{2}$.  The free propagator of the massive
gauge field $A_{\mu}$ is
\begin{eqnarray}
&&\VEV{0\vert T \left(A_{\mu}(x)A_{\nu}(y)\right)\vert
0}= \\
&&\qquad\qquad \frac {i}{(2\pi)^{4}}\int d^{4}k \frac
{K_{\mu\nu}(k)}{ (k^{2}-M^{2}+i\epsilon)} \, e^{-i \, k\cdot
(x-y)}.\nonumber
\end{eqnarray}
It does not have the bad high energy behavior found in the (Proca)
propagator in the unitary gauge formulation, where the would-be
Nambu-Goldstone boson is gauged away.

In the limit of vanishing mass of the vector boson, the gauge
field propagator goes over to the doubly transverse gauge,
($n^{\mu}\, D_{\mu\nu}(k)=k^{\mu}\, D_{\mu\nu}(k)=0$), the
propagator found \cite{Srivastava:2000cf} in QCD.  The numerator
of the gauge propagator $K_{\mu\nu}(k)$ also has important
simplifying properties, similar to the ones associated with the
projector $D_{\mu\nu}(k)$.  The transverse polarization vectors
for massive or massless vector boson may be taken to be
$E^{\mu}_{(\perp)}(k)\equiv -D^{\mu}_{\perp}(k),$ whereas the
non-transverse third one in the massive case is found to be
parallel to the LC gauge direction $\, E^{(3)}_{\mu}(k)=
-(M/k^{+})\, n_{\mu}$.  Its projection along the direction
transverse to $k_{\mu}$ shares the spacelike vector property
carried by $E^{\mu}_{(\perp)}(k)$.  The Goldstone boson or
electroweak equivalence theorem becomes transparent in  the LF
formulation.

The interaction Hamiltonian of the Standard Model can be written
in a compact form by retaining the dependent components $A^{-}$
and $\psi_{-}$ in the formulation.  Its form closely resembles the
interaction Hamiltonian of covariant theory, except for the
presence of additional instantaneous four-point interactions.  The
resulting Dyson-Wick perturbation theory expansion based on
equal-LF-time ordering has also been constructed, allowing one to
perform higher-order computations in a straightforward fashion.
The singularities in the noncovariant pieces of the field
propagators may be defined using the causal ML prescription for
$1/k^{+}$ when we employ dimensional regularization, as was shown
in our earlier work on QCD.  The power-counting rules in LC gauge
then become similar to those found in covariant gauge theory.

Spontaneous symmetry breaking is thus implemented in a novel way
when one quantizes the Standard Model at fixed light-front time
$\tau = x^+.$ In the general case, the Higgs field $\phi_i(x)$ can
be separated into two components:
\begin{equation} \phi_i(\tau,x^-,\vec x_\perp) = \omega_i(\tau,\vec x_\perp)
+ \varphi(\tau,x^-,\vec x_\perp),
\end{equation}
where $\omega_i$ is a classical $k^+=0$   zero-mode field and
$\varphi$ is the dynamical quantized field.   Here $i$ is the
weak-isospin index.  The zero-mode component is determined  by
solving the Euler-Lagrange tree-level condition:
\begin{equation}
V_i^\prime(\omega) - \partial_\perp
\partial_\perp \omega_i= 0.
\end{equation}
A nonzero value for $\omega_i$ corresponds to spontaneous symmetry
breaking.  The nonzero $\omega_i$ couples to the  gauge boson and
Fermi fields through the Yukawa interactions of the Standard
Model.  It can then  be eliminated from the theory in favor of
mass terms for the fundamental matter fields in the effective
theory.  The resulting masses are identical to those of the usual
Higgs implementation of spontaneous symmetry breaking in the
Standard Model.

The generators of isospin rotations are defined from the dynamical
Higgs fields:
\begin{equation}
G_a = -i\int dx^\perp dx^- (\partial_- \varphi)_i
(t_a)_{ij}\varphi_j \ .
\end{equation}
Note that the weak-isospin charges and the currents corresponding
to $G_a$  are not conserved if the zero mode $\omega_i$ is nonzero
since the cross terms in $\varphi,$ and $\omega$ are missing. Thus
$[H_{LF},G_a] \ne 0.$ Nevertheless, the charges annihilate the
vacuum: $G_a\ket 0_{LF}=0,$ since the dynamical fields $\varphi_i$
have no support on the LF vacuum, and all quanta have positive
$k^+.$ Thus the LF vacuum remains equal to the perturbative
vacuum; it is unaffected by the occurrence of spontaneous symmetry
breaking.

In effect one can interpret the $k^+=0$ zero mode field $\omega_i$
as an $x^-$-independent external field, analogous to an applied
constant electric or magnetic field in atomic physics.  In this
interpretation, the zero mode is  a remnant of a Higgs field which
persists from early cosmology; the LF vacuum however remains
unchanged and unbroken.

\section{Non-Perturbative Methods}

As noted in section 2., solving a quantum field theory at fixed
light-front time $\tau$ can be formulated as a relativistic
extension of Heisenberg's matrix mechanics.  If one imposes
periodic boundary conditions in $x^- = t + z/c$, then the $+$
momenta become discrete: $k^+_i = \frac{2\pi}{L} n_i, P^+ =
\frac{2\pi}{L} K$, where $\sum_i n_i =
K$.\cite{Maskawa:1975ky,Pauli:1985pv}  For a given ``harmonic
resolution" $K$, there are only a finite number of ways a set of
positive integers $n_i$ can sum to a positive integer $K$. Thus at
a given $K$, the dimension of the resulting light-front Fock state
representation of the bound state is rendered finite without
violating Lorentz invariance.  The eigensolutions of a quantum
field theory, both the bound states and continuum solutions, can
then be found by numerically diagonalizing a frame-independent
light-front Hamiltonian $H_{LC}$ on a finite and discrete
momentum-space Fock basis.   The continuum limit is reached for $K
\to \infty.$ This formulation of the non-perturbative light-front
quantization problem is called ``discretized light-cone
quantization" (DLCQ).\cite{Pauli:1985pv} The method preserves the
frame-independence of the Front form.

The DLCQ method has been used extensively for solving one-space
and one-time theories\cite{Brodsky:1997de}, including applications
to supersymmetric quantum field theories\cite{Matsumura:1995kw}
and specific tests of the Maldacena
conjecture.\cite{Hiller:2001mh}   There has been progress in
systematically developing the computation and renormalization
methods needed to make DLCQ viable for QCD in physical spacetime.
For example, John Hiller, Gary McCartor, and
I~\cite{Brodsky:2001ja,Brodsky:2001tp,Brodsky:2002tp} have shown
how DLCQ can be used to solve 3+1 theories despite the large
numbers of degrees of freedom needed to enumerate the Fock basis.
A key feature of our work is the introduction of Pauli Villars
fields to regulate the UV divergences and perform renormalization
while preserving the frame-independence of the theory.  A recent
application of DLCQ to a 3+1 quantum field theory with Yukawa
interactions is given in the references.\cite{Brodsky:2001ja} One
can also define a truncated theory by eliminating the higher Fock
states in favor of an effective
potential.\cite{Pauli:2001vi,Pauli:2001np,Frederico:2002vs}
Spontaneous symmetry breaking and other nonperturbative effects
associated with the instant-time vacuum are hidden in dynamical or
constrained zero modes on the light-front.  An introduction is
given by McCartor and Yamawaki ~\cite{McCartor:hj,Yamawaki:1998cy}

The pion distribution amplitude has been computed  using a
combination of the discretized DLCQ method for the $x^-$ and $x^+$
light-front coordinates with a spatial lattice
\cite{Bardeen:1979xx,Dalley:2001gj,Dalley:1998bj,Burkardt:2001mf}
in the transverse directions.  A finite lattice spacing $a$ can be
used by choosing the parameters of the effective theory in a
region of renormalization group stability to respect the required
gauge, Poincar\'e, chiral, and continuum symmetries.
Dyson-Schwinger models \cite{Hecht:2000xa} can also be used to predict
light-front wavefunctions and hadron distribution amplitudes by
integrating over the relative $k^-$ momentum of the Bethe-Salpeter
wavefunctions.  Explicit nonperturbative light-front wavefunctions have
been found in this way for the Wick-Cutkosky model, including spin-two
states.\cite{Karmanov:ck}
One can also implement
variational methods, using the structure of perturbative solutions as a
template for the numerator of the light-front wavefunctions.

\section{A Light-Front Event Amplitude Generator}

The light-front formalism can be used as an ``event amplitude
generator" for high energy physics reactions where each particle's
final state is completely labelled in momentum, helicity, and
phase.  The application of the light-front time evolution operator
$P^-$ to an initial state systematically generates the tree and
virtual loop graphs of the $T$-matrix in light-front time-ordered
perturbation theory in light-cone gauge.  Given the interactions
of the light-front interaction Hamiltonian, any amplitude in QCD
and the electroweak theory can be computed.  For example, this
method can  be used to automatically compute the hard-scattering
amplitudes $T_H$ for the deuteron form factor or $pp$ elastic
scattering.

At higher orders, loop integrals only involve integrations over
the momenta of physical quanta and physical phase space $\prod
d^2k_{\perp i} d k^+_i$.  Renormalized amplitudes can be
explicitly constructed by subtracting from the divergent loops
amplitudes with nearly identical integrands corresponding to the
contribution of the relevant mass and coupling counter terms --
the ``alternating denominator method".\cite{Brodsky:1973kb}  The
natural renormalization scheme to use for defining the coupling in
the event amplitude generator is a physical effective charge such
as the pinch scheme.\cite{Cornwall:1989gv}  The argument of the
coupling is then unambiguous.\cite{Brodsky:1994eh}  The DLCQ
boundary conditions can be used to discretize the phase space and
limit the number of contributing intermediate states without
violating Lorentz invariance.  Since one avoids dimensional
regularization and nonphysical ghost degrees of freedom, this
method of generating events at the amplitude level could provide a
simple but powerful tool for simulating events both in QCD and the
Standard Model.  Alternatively, one can construct the $T-$matrix
for scattering in QCD using light-front quantization and the event
amplitude generator; one can then probe its spectrum  by finding
zeros of the resolvant.  It would be particularly interesting to
apply this method to finding the gluonium spectrum of QCD.

\section*{Acknowledgments}
Work supported by the Department of Energy under contract number
DE-AC03-76SF00515.  This talk is in large part based on
collaborations with the late Professor Prem Srivastava.  I
am grateful to Professor Koichi Yamawaki and the other organizers of
this meeting for their outstanding hospitality in Nagoya.

\end{document}